\documentclass[conference,letterpaper]{IEEEtran}

\usepackage[utf8]{inputenc} 
\usepackage[T1]{fontenc}
\usepackage{ifthen}
\usepackage{cite}
\usepackage{dsfont}
\usepackage{soul}
\usepackage{comment}
\usepackage[cmex10]{amsmath}
\usepackage{url} 

\usepackage{epsfig,rotating,setspace,latexsym,epsf,amssymb,amsfonts,bm,theorem,subfigure,epstopdf}
\usepackage{bbm}

\usepackage{color}
\usepackage{mathtools}

\newtheorem{theorem}{Theorem}

\newtheorem{lemma}{Lemma}
\newtheorem{corollary}{Corollary}
\newenvironment{Proof}[1]{\medskip\par\noindent{\bf Proof:\,}\,#1}{{\mbox{\,$\blacksquare$}\par}}
\IEEEoverridecommandlockouts
\allowdisplaybreaks

\mathtoolsset{showonlyrefs}

\begin{document}
\title{Age of $k$-out-of-$n$ Systems on a Gossip Network}


\author{%
 \IEEEauthorblockN{Erkan Bayram,~ Melih Bastopcu,~ Mohamed-Ali Belabbas,~ Tamer Başar}
 \IEEEauthorblockA{Coordinated Science Laboratory\\
 University of Illinois at Urbana-Champaign,            Urbana, IL, 61801, USA\\
                   Email: \{ebayram2,bastopcu,belabbas,basar1\}@illinois.edu}
                   \thanks{Research was supported in part by ARO MURI Grant AG285, NSF-CCF 2106358, ARO W911NF-24-1-0105 and AFOSR FA9550-20-1-0333.}
}



\maketitle

\begin{abstract}
   We consider information update systems on a gossip network, which consists of a single source and $n$ receiver nodes. The source encrypts the information into $n$ distinct keys with version stamps, sending a unique key to each node. For decoding the information in a $k$-out-of-$n$ system, each receiver node requires at least $k\!+\!1$ different keys with the same version, shared over peer-to-peer connections. Each node determines $k$ based on a given function, ensuring that as $k$ increases, the precision of the decoded information also increases. We consider two different schemes: a memory scheme (in which the nodes keep the source's current and previous encrypted messages) and a memoryless scheme (in which the nodes are allowed to only keep the source's current message). We measure the ``timeliness'' of information updates by using the $k$-keys version age of information. Our work focuses on determining closed-form expressions for the time average age of information in a heterogeneous random graph under both with memory and memoryless schemes.   
\end{abstract}
\vspace{-1mm}
\section{Introduction}
\vspace{-1mm}
In a peer-to-peer sensor or communication network, information spreading can be categorized into single-piece dissemination (when one node shares its information) and multicast dissemination (when all nodes share their information)~\cite{bayram2023vector}. But some applications lie between two categories, in which a node needs to collect a multitude of messages or observations generated at the same time to construct meaningful information. If any $k+1$ out of a total of $n$ messages are sufficient to construct the information in a system, it is called a $k$-out-of-$n$ system. Such systems find applications in various domains including robotics, cryptography, and communication systems. 

For example, in cryptography, the information source can apply $(k,n)$-Threshold Signature Scheme (TSS)~\cite{hwang2003practical} on the information to put it into $n$ distinct keys such that any subset of $n$ keys with $k+1$ cardinality would be sufficient to decode the encrypted message. Furthermore, $k$-out-of-$n$ systems enhance error correction performance in data transmission. For instance, the relative positioning of the target object using the Time Difference of Arrival (TDoA) technique~\cite{gustafsson2003positioning} requires at least three simultaneously generated Time of Flight measurements while more measurement increase the accuracy of estimation of relative position of the object. Another common example is $(n,k+1)$-MDS error correction codes~\cite{buyukates2020timely}, in which any $k+1$ codeword suffices for message decoding, with additional redundant codewords that facilitate error correction.

Motivated by these applications, we consider in this work an information source that generates updates and then encrypts (encodes) them by using a $k$-out-of-$n$ systems, e.g. $(k,n)$-TSS. For the sake of simplicity, we consider the source using $(k,n)$-TSS, but it is worth noting that our results are applicable to {\em any $k$-out-of-$n$ system}, some of examples are discussed above. The source is able to send the encrypted messages to the $n$ receiver nodes instantaneously. Upon receiving these updates, the nodes start to share their local messages with their neighboring nodes to decrypt the source's update. Each node determines the required number of keys $k$ to achieve a given precision rate $\alpha$ on the information for a given precision-rate function $D(k,n,\beta)$, which will be introduced later. The nodes that get $k$ different messages of the same update from their neighbors can decode source's information. We study two different settings where  ($i$) the nodes have memory, in which case they can hold the current and also the previous keys received from the source, and ($ii$) the nodes do not have memory, in which case they can only hold the keys from the most current update. For both of these settings, we study the information freshness achieved by the receivers as a result of applying $(k,n)$-TSS. Age of information (AoI) has been introduced as a new performance metric to measure the freshness of information in communication networks \cite{Kaul12a}. Inspired by recent studies, advancements in Age of Information (AoI) have been made in various gossip network scenarios, including those examining scalability, optimization, and security~\cite{yates2021age, Buyukates2022, Bastopcu2021gossip, Bastopcu2024gossip, mitra2023ageaware, delfani2023version, Kaswan2024timestomping, kaswan2023choosing}. In all these aforementioned works, the source sends its information to gossip nodes without using any encryption. 

For the first time, in this work, we consider the version age of information in a gossip network where the source encrypts the information. Then, we have the following contributions:
\vspace{-1.5mm}
\begin{itemize}
    \item We derive closed-form expressions for the time average of $k$-keys version age for an arbitrary non-homogeneous network (e.g., independently activated channels). 
\item We show that the time average of the $k$-keys version age for a node in both schemes decreases as the edge activation rate increases, or as the number of keys required to decode the information decreases, or as the number of gossip pairs in the network increases in the numerical results, 
\item We show that a memory scheme yields a lower time average of $k$-keys version age compared to a memoryless scheme. However, the difference between the two schemes diminishes with infrequent source updates, frequent gossip between nodes, or a decrease in $k$ for a fixed number of nodes. 
\end{itemize}

\section{System Model and Metric}
We consider an information updating system consisting of a single source, which is labeled as node $0$, and $n$ receiver nodes. The information at the source is updated at times distributed according to a Poisson counter, denoted by $N_0(t)$, with  rate $\lambda_s$. We refer to the time interval between $\ell$th and $(\ell+1)$th information updates (messages) as the {\em $\ell$th version cycle} and denote it by $U^\ell$. Each update is stamped by the current value of the process $N_0(t)=\ell$ and the time of the  $\ell$th update is labelled $\tau_\ell$  once it is generated. The stamp $\ell$ is called {\em version-stamp} of the information.

We assume that the source is able to instantaneously encrypt the information update by using $(k,n)$-TSS once it is generated. To be more precise, we assume that the source puts the information update into $n$ distinct keys and sends one of the unique keys to each receiver node at the time $\tau_\ell$, instantaneously. Once a node gets a unique key from the source at $\tau_\ell$ for version $\ell$, it is aware of that there is new information at the source. Each node wishes its knowledge of the source to be as timely as possible. The timeliness is measured for an arbitrary node $j$ by the difference between the latest version of the message at the source node, $N_0(t)$, and the latest version of the message which can be {\em decrypted} at node $j$, denoted by $N^k_j(t)$. This metric has been introduced as {\em version age of information} in~\cite{yates2021age, Eryilmaz2021}. We call it  {\em $k$-keys version age of node $j$} at time $t$ and denote it as \begin{align}\label{eqn:defn_process}
\\[-1.5em]
    A^k_j(t) := N_0(t) - N^k_j(t).\\[-1.5em]
\end{align}
Recall that in the $(k,n)$-TSS, a node needs to have $k+1$ keys with the version stamp $\ell$ in order to decrypt the information at the source generated at $\tau_\ell$. Since the source sends a unique key to all receiver nodes, a node needs $k$ additional distinct keys with version $\ell$ to decrypt the $\ell$th message. 
We denote by $\Vec{G}=(V,\Vec{E})$ the directed graph with node set $V$ and edge set $\Vec{E}$. We let  $\Vec{G}$ represent the communication network according to which nodes exchange information. If there is a directed edge $e_{ij}\in\vec{E}$, we call node $j$ {\em gossiping neighbor} of node $i$. We consider a precision-rate function given by $D(k,n,\beta):\mathbb{N}\times\mathbb{N}\times\mathbb{R}^+ \to [0,1]$ that quantifies how precisely a node decodes the status update if it collects $k+1$ out of $n$ symbols, where $\beta$ represents a relevant system parameter. For a given precision rate $\alpha$, the required number of keys for node $j$ is defined as $k_j(\beta,\alpha)\!:=\!\inf\{ k \in \mathbb{N} : D(k,n,\beta)\!\geq\!\alpha\}$. For simplicity, we assume $D(k,n,\beta)\!:=\!\sum_{i=0}^{\lfloor k \rfloor} {n \choose i}\beta^{i}(1\!-\!\beta)^{n-i}$. However, any function that is monotonically increasing in $k\leq n$ for fixed $n$ can serve as a precision-rate function. Here, the rate $\beta$ represents the amount of noise in each keys. We call a communication network $(k,n)$-TSS feasible for rate $(\beta,\alpha)$ if the node $0$ has out-bound connections to all other nodes and the smallest in-degree of the receiver nodes is greater than the smallest $k_j(\beta,\alpha)$ among all nodes.\footnote{In this work, all the nodes may have different $k_j(\beta,\alpha)$s. Our results are directly applicable for any arbitrary selection of $k_j(\beta,\alpha)$s. For that, we may omit the user index $j$ and instead use $k$, directly.}

We consider a $(k,n)$-TSS feasible network, in which, nodes are allowed to communicate and share {\em only} the keys that are received from the source with their {\em gossiping neighbor}. The edge $e_{ij}$ is activated at times distributed according to the Poisson counter $N^{ij}(t)$, which has a rate $\lambda_{ij}$ and once activated, node $i$ sends a message to node $j$, instantaneously. All counters are pairwise independent. This process occurs under two distinct schemes: {\em with memory} and {\em memoryless}.

\begin{figure}[t]
    \centering
    \includegraphics[width=1\linewidth]{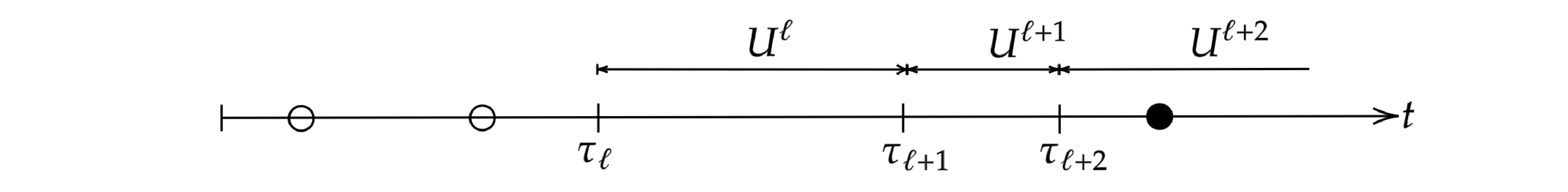}
    \vspace{-0.55cm}
     \caption{Sample timeline of the source update and the edge $e_{ij}$ activation. The last activation of $e_{ij}$ is marked by ($\bullet)$ and the previous activations of $e_{ij}$ are marked by ($\circ$).}\label{fig:timeline_for_schemes}
          \vspace{-0.5cm}
          \end{figure}

In the memory scheme, nodes can store (and send) the keys of the previous updates. For example, if the edge $e_{ij}$ is activated at $t$, node $i$ sends node $j$ {\em all} the keys that the source has sent to node $i$ since the last activation of $N^{ij}(t)$ before $t$. For the illustration in Fig.~\ref{fig:timeline_for_schemes}, node $i$ sends the set of keys with the versions $\{\ell,\ell+1,\ell+2\}$ to node $j$ in the memory scheme. Note that this can be implemented by finite memory in a finite node network with probability $1$; we will provide below the distribution of the number of keys in the message. In the memoryless scheme, nodes have no memory and only store the latest key obtained from the source. If the edge $e_{ij}$ is activated at $t$, node $i$ sends node $j$ {\em only} the last key that the source sent to node $i$ before $t$. Referring again to the illustration in Fig.~\ref{fig:timeline_for_schemes}, node $i$ in this case sends only the key with the version $\{\ell+2\}$ to node $j$.

\begin{figure}[t]
	\begin{center}
		\subfigure[]{%
			\includegraphics[width=0.72\linewidth]{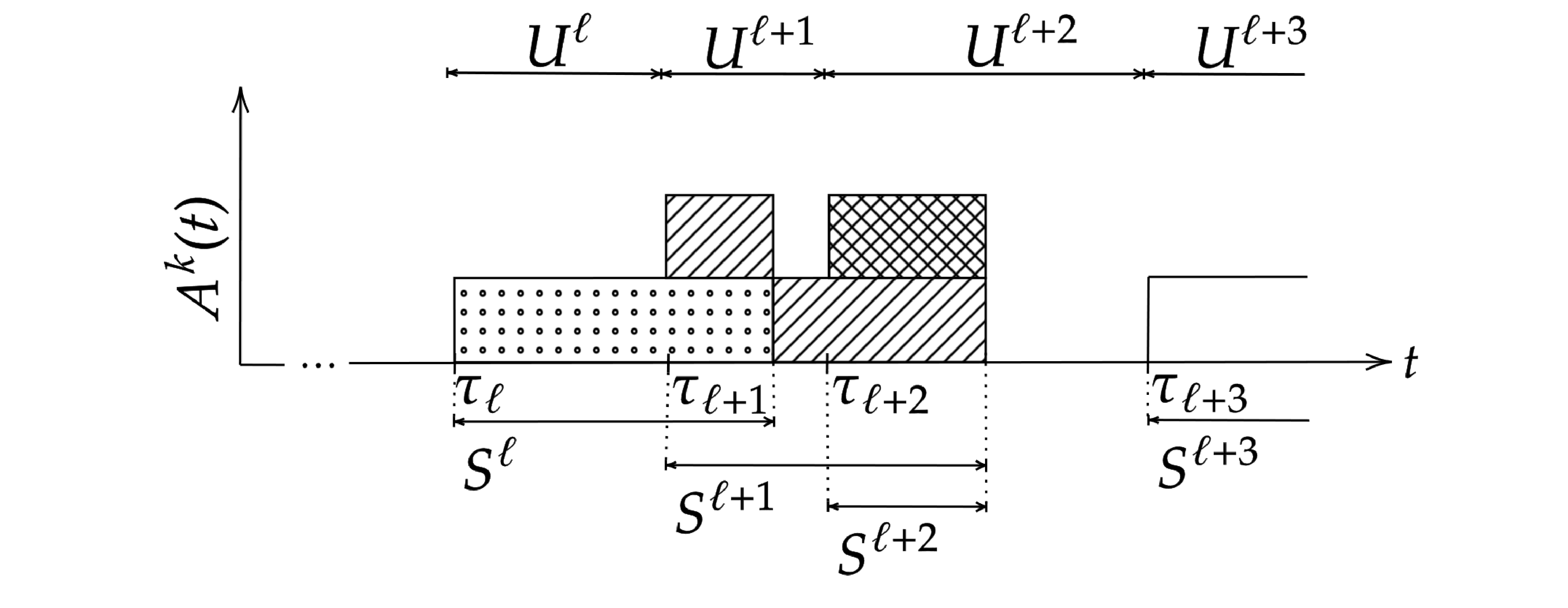}} 
		\subfigure[]{%
			\includegraphics[width=0.72\linewidth]{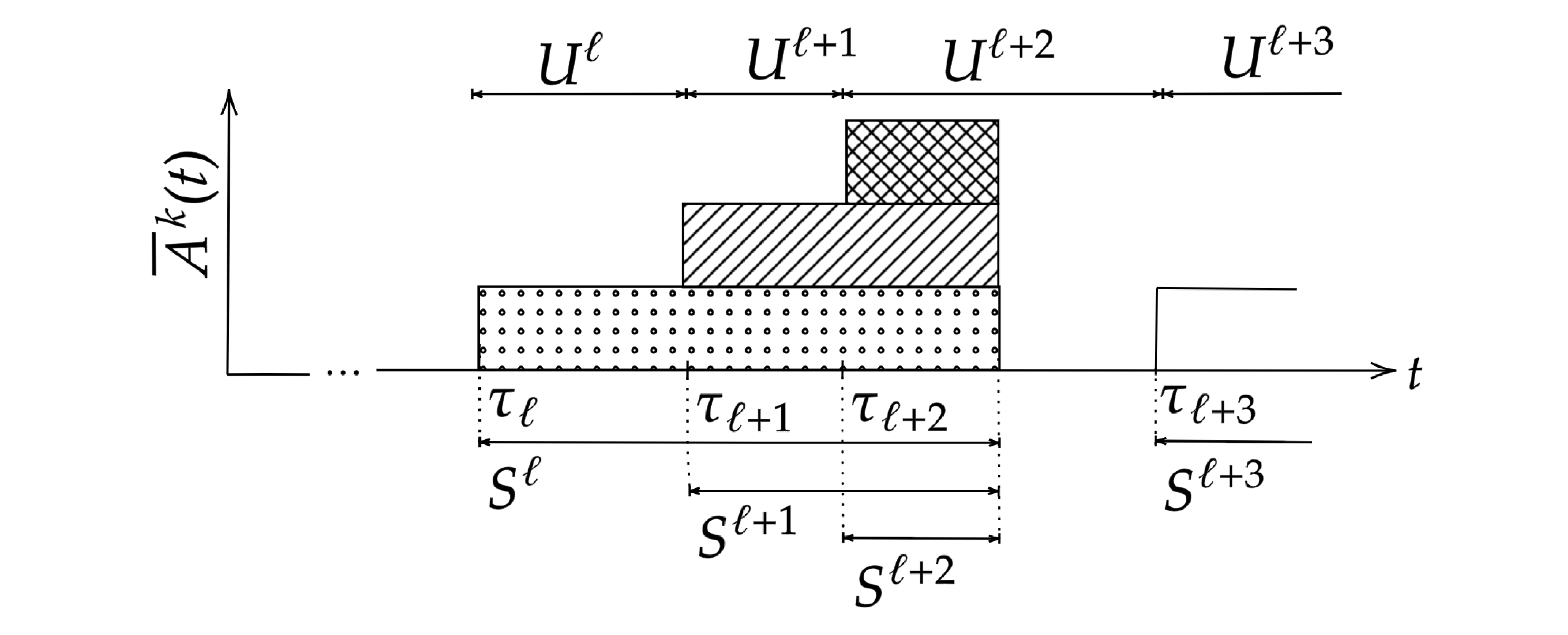}}
	\end{center}
 \vspace{-5mm}
	\caption{Sample path of the $k$-keys version age (a) $A^k(t)$ for a node with memory and (b) $\bar{A}^k(t)$ for a node without memory.}
	\label{fig:path_w_memory}
	\vspace{-6mm}
\end{figure} 

Fig.~\ref{fig:path_w_memory}(a) and Fig.~\ref{fig:path_w_memory}(b) depict the sample path of the $k$-keys version age process ${A}^k(t)$ (resp.~ $\bar{A}^k(t)$) for a node with memory (resp.~without memory). It is worth noting that we associate the notation $\bar{\cdot}$ with the memoryless scheme. It is assumed that edge activations and source updates occur at the same time in both schemes in the figures. In the memory scheme, we define {\em the service time of information with version $\ell$ to an arbitrary node $j$}, denoted by $S^\ell_j$, as the duration between $\tau_\ell$ and the time when node $j$ can decrypt the information with version $\ell$, as shown in Fig.~\ref{fig:path_w_memory}(a). In the memoryless scheme, a node can miss an information update with version $\ell$ if it cannot get $k$ more distinct keys before the next update arrives at $\tau_{\ell+1}$. Thus, for a node without memory, we define $S^\ell_j$ as the duration between $\tau_\ell$ and the earliest time when the node can decrypt information with a version of at least $\ell$. In Fig.~\ref{fig:path_w_memory}(b), the node could only decode the information with version $\ell\!+\!2$ while missing $\ell$ and $\ell\!+\!1$. Therefore, the service times $S^\ell$ and $S^{\ell+1}$ end at the same time as the service time $S^{\ell+2}$ ends. 

Let $\Delta^k_j(t)$ be the total $k$-keys version age of node $j$, defined as the integrated $k$-keys version age of node $j$, $A^k_j(\tau)$, until time $t$. For both schemes, the time average of $k$-keys version age process of node $j$ is defined as follows:
\begin{align*}
    \\[-45pt]
\end{align*}
\begin{align}\label{eqn:defn_time_av}
    \Delta^{k}_j :=\lim_{t\to \infty} \frac{\Delta^{k}_j (t)}{t}  = \lim_{t\to \infty} \frac{1}{t} \int_{0}^{t} A^k_j(\tau) d\tau.
\end{align}
\begin{align*}
    \\[-45pt]
\end{align*}
We interchangeably call $\Delta^{k}_j$ {\em the version age of $k$-keys for node $j$} on $({\beta},\alpha)$. If nodes in the network have no memory, we denote {\em the version age of $k$-keys for node $j$} by $\bar{\Delta}^k_j$. 

\section{Age Analysis}
In this section, we first introduce the main concepts that will be useful to derive age expressions and then provide closed-form expressions for the version age of $k$-keys with memory, $\Delta^{k}_j$; and without memory, $\bar{\Delta}^{k}_j$.

Consider a set of random variables $\mathcal{Y}=\{Y_i\}_{i=1}^n$. We denote the $k^{th}$ smallest variable in the set $\mathcal{Y}$ by $Y_{(k:n)}$. We call $Y_{(k:n)}$ the $k$th order statistic of $n$-samples ($k=1,2,\cdots,n$) in the set $\mathcal{Y}$. Let $\mathcal{N}^+_j=\{ i\in V , e_{ij} \in \Vec{E} \}$ be the set of nodes with out-bound connections to node $j$; denote its cardinality by $n_j$. Let $X_{ij}$ be the times between successive activations of the edge $e_{ij}$. Then, $X_{ij}$ is an exponential random variable with mean $1/\lambda_{ij}$. Let $\mathcal{X}_j$ be the set of random variables $X_{ij}$, $\forall i \in \mathcal{N}^+_j$.
We denote the $k$th order statistic of the set $\mathcal{X}_j$ by $\mathcal{X}_{(k:n_j)}$.

\subsection{Nodes with Memory}
In this section, we provide the closed-form expression ${\Delta}^k_j$ in a $(k,n)$-TSS feasible network for a given rate $(\beta,\alpha)$. 
\vspace{-2mm}
\begin{theorem}\label{thm:hetero_w_memory_age}
Let the precision rate function $D(\cdot,\cdot,\beta)$ be given and assume that $\vec{G}$ is a $(k,n)$-TSS feasible network for a given $(\beta,\alpha)$. Consider an arbitrary node $j$ in $\Vec{G}$. The version age of $k$-keys for node $j$ with {\em memory} (at $k=k_j(\beta,\alpha)$) obeys
\begin{equation}
    \Delta^{k}_j = \frac{\mathbb{E}[\mathcal{X}_{(k:{n}_j)}]}{\mathbb{E}[U]} \mbox{ w.p. } 1.
\end{equation}
where $U$ is the interarrival time for the source update.
\end{theorem}

{It is worth noting that Theorem~\ref{thm:hetero_w_memory_age} holds for any $(k,n)$-TSS feasible network with heterogeneous (possibly different) edge activation rates. One can easily obtain an explicit form of $\Delta^k$ for a given node $j$ and $\Vec{G}$ by using the c.d.f. of order statics provided in~\cite{david2004order}. We need the following lemma for the proof of Theorem~\ref{thm:hetero_w_memory_age}.}
\vspace{-3mm}
\begin{lemma}\label{lem:service}
If nodes have memory, then the service time of the information with version $\ell$ to node $j$ is the $k$th order statistic of the set of exponential random variables $\mathcal{X}_j$.   
\end{lemma}
\vspace{-3mm}
\begin{Proof} We denote the set of the first activation times of edges that are connected to the node after $\tau_\ell$ by $\mathcal{X}^\ell$. For the case $\mathcal{X}^\ell_{(k:n_j)} \leq U^\ell$, the result trivially follows from the definitions. Consider the case $\mathcal{X}^\ell_{(k:n_j)} > U^\ell$. By definition, a new status update arrives at all nodes at $\tau_{\ell+1} (=\tau_\ell +  U^\ell$). However, the structure of a message sent after $\tau_{\ell+1}$ from a node to another node ensures that it has the key with version stamp $\ell$. Therefore, the service time for a status update is also $\mathcal{X}^\ell_{(k:n_j)}$ (regardless of the fact that a new update arrived).    
\end{Proof}

We are now in a position to
{prove Theorem~\ref{thm:hetero_w_memory_age}}. 

\begin{Proof}[Proof of Theorem~\ref{thm:hetero_w_memory_age}] Let $k$ be ${k_j(\beta,\alpha)}$ for given $(\beta,\alpha)$. Let $T:=\{\tau_\ell\}_{\ell=0}^{\infty}$ be the monotonically increasing sequence of times when status updates occur at the source node $0$, with $\tau_0\!:=\!0$. Let $T^1\!=\!\{\tau_{\ell_a}\}_{a=0}^{\infty}$ be a subsequence of $T$ such that $A^k(\tau_{\ell_a})\!=\!1$. Let $L_a$ be the time elapsed between two consecutive successful arrivals of the subsequence $T^1$. Let $R_a$ be the version age of $k$-keys, $A^k(t)$ integrated over the duration $[\tau_{\ell_a},\tau_{\ell_{a+1}})$ in a node. Then, we have:
\begin{align*}
    \frac{\mathbb{E}[R_a]}{\mathbb{E}[L_a]} \!= \frac{\mathbb{E}[\sum_{i=\ell_a}^{\ell_{a+1}-1} S^i_j  ]}{\mathbb{E}[ \sum_{i=\ell_a}^{\ell_{a+1}-1} U^i]} \!=  \frac{ \mathbb{E}[\sum_{i=\ell_a}^{\ell_{a+1}-1}\mathcal{X}^i_{(k:n_j)}  ]}{  \mathbb{E}[\sum_{i=\ell_a}^{\ell_{a+1}-1} U^i]} \!=  \frac{ \mathbb{E}[{\mathcal{X}_{(k:n_j)}}  ]}{  \mathbb{E}[U]}
\end{align*}
From Lemma~\ref{lem:service}, we have ${S}^i_j=\mathcal{X}^i_{(k:n_j)}$ for any $i$. It is worth noting that two random variables $\mathcal{X}^{i}_{(k:n_j)}$ and $\mathcal{X}^{i+1}_{(k:n_j)}$ are not independent in the memory scheme if $\ell_a\!\leq\!i\!\leq\!\ell_{a+1}-1$ for any $a \in \mathbb{N}$, but they are identically distributed. Thus, we have $\mathbb{E}[\mathcal{X}_{(k:n_j)}]=\mathbb{E}[\mathcal{X}^i_{(k:n_j)}]$ and $\mathbb{E}[U]=\mathbb{E}[U^i]$ for any $i$. By construction of the sequence $T^1$, a pair of $((L_a,R_a),(L_b,R_b))$ for any $a \neq b$ is $i.i.d.$.  Then, from~\cite[Thm. 6]{gallager1997discrete}, we find the time average $\Delta^k_j$, which completes the proof.
\end{Proof}    
The case of a fully connected directed graph $\vec{G}$ on $n+1$ nodes (including the source node) with $\lambda_{ij}={\lambda_e}/{(n-1)}$ for all edges $e_{ij}$ in $\vec{E}$,  is called  {\em scalable homogeneous network} (SHN) and $\lambda_e$ is the {\em gossip rate}. 
\vspace{-3mm}
\begin{corollary}\label{cor:w_memory_age}
For a SHN, the version age of $k$-keys for a node with {\em memory} is: \begin{align}\label{eqn:w_memory_age}
    \Delta^{k} = \frac{\mathbb{E}[\mathcal{X}_{(k:n-1)}]}{\mathbb{E}[U]} = \frac{(n-1)\lambda_s}{\lambda_e}  ( \sum^{n-1}_{i=n-k} \frac{1}{i} ) \mbox{ w.p. } 1.    
\end{align}
\end{corollary}
\vspace{-4mm}
{
\begin{Proof} In a SHN, every node has $n-1$ outbound connection; thus, $n_j=n-1$ and the processes $A^k_j(t)$ are statistically identical for any node $j$. Then, the set $\mathcal{X}_j$ is the set of $i.i.d$ exponential random variables with rate $\frac{\lambda_e}{(n-1)}$ for $j \in V$. From \cite{david2004order}, we have $\mathbb{E}[\mathcal{X}_{(k:n-1)}]=\frac{(n-1)}{\lambda_e} ( H_{n-1} - H_{n-1-k} )$ where $H_n= \sum_{i=1}^n \frac{1}{i}$. From Theorem.~\ref{thm:hetero_w_memory_age}, the results follows. \end{Proof}}
\vspace{-3mm}
\begin{corollary}\label{cor:w_memory_scale}
For a finite $k$ and a SHN with a countable memory, we have the following scalability result:
\begin{align*}
    \lim_{n \to \infty} \Delta^k = \frac{k\lambda_s}{\lambda_e} \mbox{ w.p. } 1.
\end{align*}
\end{corollary}
\vspace{-2mm}
One can easily take the limit as $n$ goes to $\infty$ in~\eqref{eqn:w_memory_age} to have Corollary~\ref{cor:w_memory_scale} of Theorem~\ref{thm:hetero_w_memory_age}. We now elaborate on the {\em number of keys} in the message that is sent over an edge. Let $\mathcal{M}^i_j$ be the number of keys in the message that is sent over the edge $e_{ij}$. In each update cycle, either the source is updated before the edge $e_{ij}$ is activated, which increases $\mathcal{M}^i_j$ by $1$ or the edge is activated before a source update in which case node $i$ sends all the keys to node $j$, and thus $\mathcal{M}^i_j$ reduces to $0$. From \cite[Prob. 9.4.1]{yates2014probability}, $\mathcal{M}^i_j$ has a geometric distribution with success probability $\lambda_{e_{ij}}/(\lambda_{e_{ij}}+\lambda_{s})$.

\vspace{-1mm}
\subsection{Nodes without Memory}
\vspace{-1mm}
In the following section, we analyze the memoryless scheme, in which a message has only $1$ key in any time. We provide a closed-form expression $\bar{\Delta}^k_j$, in a $(k,n)$-TSS feasible network for a given rate $(\beta,\alpha)$. 
 \vspace{-3mm}
\begin{theorem}\label{thm:hetero_wo_memory_age}
Let the precision rate function $D(\cdot,\cdot,\beta)$ be given and assume that $\vec{G}$ is a $(k,n)$-TSS feasible network for a given $(\beta,\alpha)$. Consider an arbitrary node $j$ in $\Vec{G}$. The version age of $k$-keys for node $j$ {\em without  memory} (at $k=k_j(\beta,\alpha)$) obeys
\begin{equation}\label{eqn:hetero_wo_memory_age}
    \bar{\Delta}^{k}_j = \frac{\mathbb{E}[\min( \mathcal{X}_{(k:{n}_j)}, U )]}{Pr(\mathcal{X}_{(k:n_j)} \leq U ) \mathbb{E}[U]} \mbox{ w.p. } 1.
\end{equation}
where $U$ is the interarrival time for the source update. 
\end{theorem}

\vspace{-2mm}
We need the following Lemma for the proof of Theorem~\ref{thm:hetero_wo_memory_age}. 

\vspace{-2.mm}
\begin{lemma}\label{lem:markov}
Let $\bar{A}^k_j[\ell]=\Bar{A}^k_j(\tau_\ell)$ be the version age of information at $\tau_\ell$ for the node $j$. The sequence $\bar{A}^k_j[\ell]$ is homogeneous success-run with the rate $\mu_j\!=\!Pr(\mathcal{X}^j_{(k:n_j)}\!\leq\!U)$.
\end{lemma}
\vspace{-2.mm}
{
\begin{Proof}
We remove the index $j$ and the number of keys $k$ on notation, as it would be sufficient to prove the results for an arbitrary node and $k>0$. We denote the set of the first arrival times of edges that are connected to the node after $\tau_\ell$ by $\mathcal{X}^\ell$. One can easily show that $(\mathcal{X}^\ell_{(k:n-1)},U^\ell)$ and $(\mathcal{X}^{\ell+1}_{(k:n-1)},U^{\ell+1})$ for any $\ell$ are independent. This implies that the sequence ${A}[\ell]$ has Markov property and it evolves as follows:
\begin{align*}
    \\[-45pt]
\end{align*}
\begin{align}
    \bar{A}[\ell+1]= \begin{cases}
        1 & \mbox{if }  \mathcal{X}^{\ell}_{(k:n-1)}\leq U^\ell \\
        \bar{A}[\ell] + 1 & \mbox{if }    \mathcal{X}^{\ell}_{(k:n-1)} > U^\ell
    \end{cases}
\end{align} 
\begin{align*}
    \\[-45pt]
\end{align*}
and $\bar{A}[0]$ is $1$ by definition. 
Here, $\bar{A}[\ell]$ is a discrete-time Markov chain on infinite states $\mathcal{A}\!:=\!\{1,2,\cdots\}$ with the initial distribution $\pi\!=\![1,0,\cdots]$ and the state transition matrix of $\!P$
$$
P := \left[ \begin{smallmatrix}
        \mu & 1-\mu &   &   \cdots \\ 
        \mu &    &  1-\mu &  \cdots \\
        \vdots & &      &    \ddots & 
    \end{smallmatrix} \right]
$$ where $\mu\!=\!Pr( \mathcal{X}^{\ell}_{(k:n-1)}\!\leq\!U^\ell)$. Then, it says the sequence $\bar{A}[\ell]$ is a homogeneous success-run chain with rate $\mu$.\end{Proof}
As an easy corollary to Lemma~\ref{lem:markov}, the random variable $\bar{A}_j[\ell]$ has truncated geometric distribution at $\ell+1$ with success rate $\mu_j$. Now, we can prove Theorem~\ref{thm:hetero_wo_memory_age}.

\begin{Proof}[Proof of Theorem~\ref{thm:hetero_wo_memory_age}] 
Let $k$ be ${k_j(\beta,\alpha)}$ for given $\beta$ and $\alpha$. Let $T^1\!=\!\{\tau_{\ell_a}\}_{a=0}^{\infty}$ be a subsequence of $T$ such that $\bar{A}^k_j(\tau_{\ell_a})\!=\!1$. Let $L_a$ be the time elapsed between two consecutive successful arrivals of the subsequence $T^1$. Let $R_a$ be the version age of $k$-keys, $\bar{A}^k_j(t)$ integrated over the duration $[\tau_{\ell_a},\tau_{\ell_{a+1}})$ in a node. Then, we have $R_a$ as follows:
\begin{align*}\\[-2.5em]
        R_a = \int_{\tau_{\ell_{a}}}^{\tau_{\ell_{a+1}}} \bar{A}^k_j(t) d t\!= \sum^{\ell_{a+1}-1}_{i=\ell_{a}} \bar{A}^k_j[i] ( \mathds{1}_{E_i} \mathcal{X}^{i}_{(k:n_j)} + \mathds{1}_{E_i^c} U^{i} ) \\[-2em]       
\end{align*}
where the event $E_i=\{\mathcal{X}^{i}_{(k:n_j)}\leq U^{i}\}$ and we denote complement by $E_i^c$. Then, we have;
\begin{align} \\[-2.1em]
\frac{\mathbb{E}[R_a]}{\mathbb{E}[L_a]} &= \frac{\mathbb{E}[\sum^{\ell_{a+1}-1}_{i=\ell_{a}} \bar{A}[i] \min(  \mathcal{X}^{i}_{(k:n_j)},U^{i} )]}{\mathbb{E}[\sum^{\ell_{a+1}-1}_{i=\ell_{a}} U^i]}\\[-2.1em]
\end{align}
 A pair of random variables $\mathcal{X}^{i}_{(k:n_j)}$ and $\mathcal{X}^{i+1}_{(k:n_j)}$ are identically distributed. Thus, we have $\mathbb{E}[\mathcal{X}_{(k:n_j)}]=\mathbb{E}[\mathcal{X}^i_{(k:n_j)}]$ and $\mathbb{E}[U]=\mathbb{E}[U^i]$ for any $i$. Let $M_a:=\bar{A}[\ell_{a+1}-1]$ be the number of information updates that the node has missed between $\tau_{\ell_a}$ and $\tau_{\ell_{a+1}}$ (that is, $M_a:=\ell_{a+1}-\ell_{a}$). Then, we have:
\begin{align}\label{eqn:r_a_computation}\\[-2em]
        \frac{\mathbb{E}[R_a]}{\mathbb{E}[L_a]}  &= \frac{ \mathbb{E}[\min(  \mathcal{X}_{(k:n_j)},U ) ] \mathbb{E}[\frac{M_a(M_a+1)}{2}] }{\mathbb{E}[U]\mathbb{E}[M_a] }\\[-2em]
\end{align}
From Lemma~\ref{lem:markov}, we know that $\bar{A}[\ell]$ is a succes-run chain with rate $\mu$, then $M_a$ has geometric distribution for sufficiently large $a \in \mathbb{N}$ with rate $\mu(=Pr( \mathcal{X}_{(k:n_j)} \leq U))$. Then, we have:
\begin{align*}\label{eqn:wo_age_comp}\\[-2em]
        \frac{\mathbb{E}[R_a]}{\mathbb{E}[L_a]} &= \frac{\mathbb{E}[\min(  \mathcal{X}_{(k:n_j)},U ) ] ( \frac{2-\mu}{2\mu^2}\!+\!\frac{1}{2\mu} ) }{\frac{1}{\mu} \mathbb{E}[U]}  =  \frac{\mathbb{E}[\min(  \mathcal{X}_{(k:n_j)},U ) ]}{\mu\mathbb{E}[U]}\\[-2em]
\end{align*}
By construction, a pair of $((L_a,R_a),(L_b,R_b))$ for any $a \neq b$ is $i.i.d.$. From~\cite[Thm. 6]{gallager1997discrete}, we find the time average $\bar{\Delta}^k_j$.\end{Proof} 

\vspace{-3mm}
\begin{corollary}\label{cor:wo_memory_age}
For a SHN, the version age of $k$-keys TSS for an individual node {\em without memory} is: 
\begin{equation}\label{eqn:cor_wo_memory}\\[-1em]
    \bar{\Delta}^k = \frac{ \frac{\lambda_s}{\lambda_e\!+\!\lambda_s}\!+\!\sum\limits_{j=2}^k\left(\frac{\lambda_s}{\lambda_e \frac{(n-j)}{(n-1)}\!+\!\lambda_s}\!\prod\limits_{i=1}^{j\!-\!1}\frac{\lambda_e \frac{(n-i)}{(n-1)}}{\lambda_e \frac{(n-i)}{(n-1)}\!+\!\lambda_s }  \right) }{Pr(\mathcal{X}_{(k:n-1)}\leq U)} \mbox{ w.p. } 1 
\end{equation}
\end{corollary}
\begin{Proof}
    Let $Y^k$ be $\min\{\mathcal{X}_{(k:n-1)},U\}$. Let $\Tilde{X}_{i:n-1}$ be the difference between $i$th and $(i\!-\!1)$th order statistics of the set $\mathcal{X}$ for $i>1$ and $\Tilde{X}_{1:n-1}\!=\!\mathcal{X}_{(1:n-1)}$. Let $\Tilde{Y}_i = \min\{\Tilde{X}_{i:n-1},U^\ell\}$. One can see that, from memoryless property, $\Tilde{X}_{i:n-1}$ corresponds to the minimum of a set of $(n\!-\!i)$ $i.i.d.$ exponential random variables (r.v.) with mean $\frac{(n-1)}{\lambda_e}$. Thus, r.v. $\Tilde{X}_{i:n-1}$ is also an exponential r.v. with the parameter $\frac{1}{\lambda_e\mathcal{B}(n,i)}$ where $\mathcal{B}(n,i)=\frac{(n-i)}{(n-1)}$. Thus, r.v. $\Tilde{Y}_i$ is the minimum of two independent exponentially distributed r.v. From~\cite[Prob. 9.4.1]{yates2014probability}, r.v. $\Tilde{Y}_i$ is exponentially distributed with the mean $\frac{1}{\lambda_e \mathcal{B}(n,i) + \lambda_s}$ 
and we have: 
\begin{align*}
    \\[-45pt]
\end{align*}
\begin{align}\label{eqn:prob_ul}\\[-2.1em]
    Pr(U^\ell\!>\!\Tilde{X}_{i:n-1})= \frac{\lambda_e \mathcal{B}(n,i)}{\lambda_e  \mathcal{B}(n,i) + \lambda_s } .\\[-2.1em]
\end{align}
 for $i\geq1$. From the total law of expectation and memoryless property of $U^\ell$, we have the following:
\begin{align}\label{eqn:exp_y_k}
    \mathbb{E}[Y^k] =& Pr( U^\ell\!\leq\!\Tilde{X}_{1:n-1} ) \mathbb{E}[\Tilde{Y}_1] \\
                    &+ \!\!Pr( U^\ell\!\leq\!\Tilde{X}_{2:n-1} )  Pr(\!U^\ell\!\!>\!\!\Tilde{X}_{1:n-1} \!)  (\mathbb{E}[\Tilde{Y}_1] \!\!+\!\! \mathbb{E}[\Tilde{Y}_2] ) \\
                    &+\ldots + \prod_{i=1}^{k-1} Pr( U^\ell\!>\!\Tilde{X}_{i:n-1} ) ( \sum_{i=1}^{k} \mathbb{E}[\Tilde{Y}_i])
 \end{align}
If we rearrange the sum above and we plug $\mathbb{E}[\Tilde{Y}_i]$ above and $Pr(U^\ell\!>\!\Tilde{X}_{i:n-1})$ in~\eqref{eqn:prob_ul} into~\eqref{eqn:exp_y_k}, we obtain Cor.~\ref{cor:wo_memory_age}. 
\end{Proof}}

\section{Numerical Results and Conclusion}
\begin{figure}
    \centering
    \includegraphics[width=0.75\linewidth]{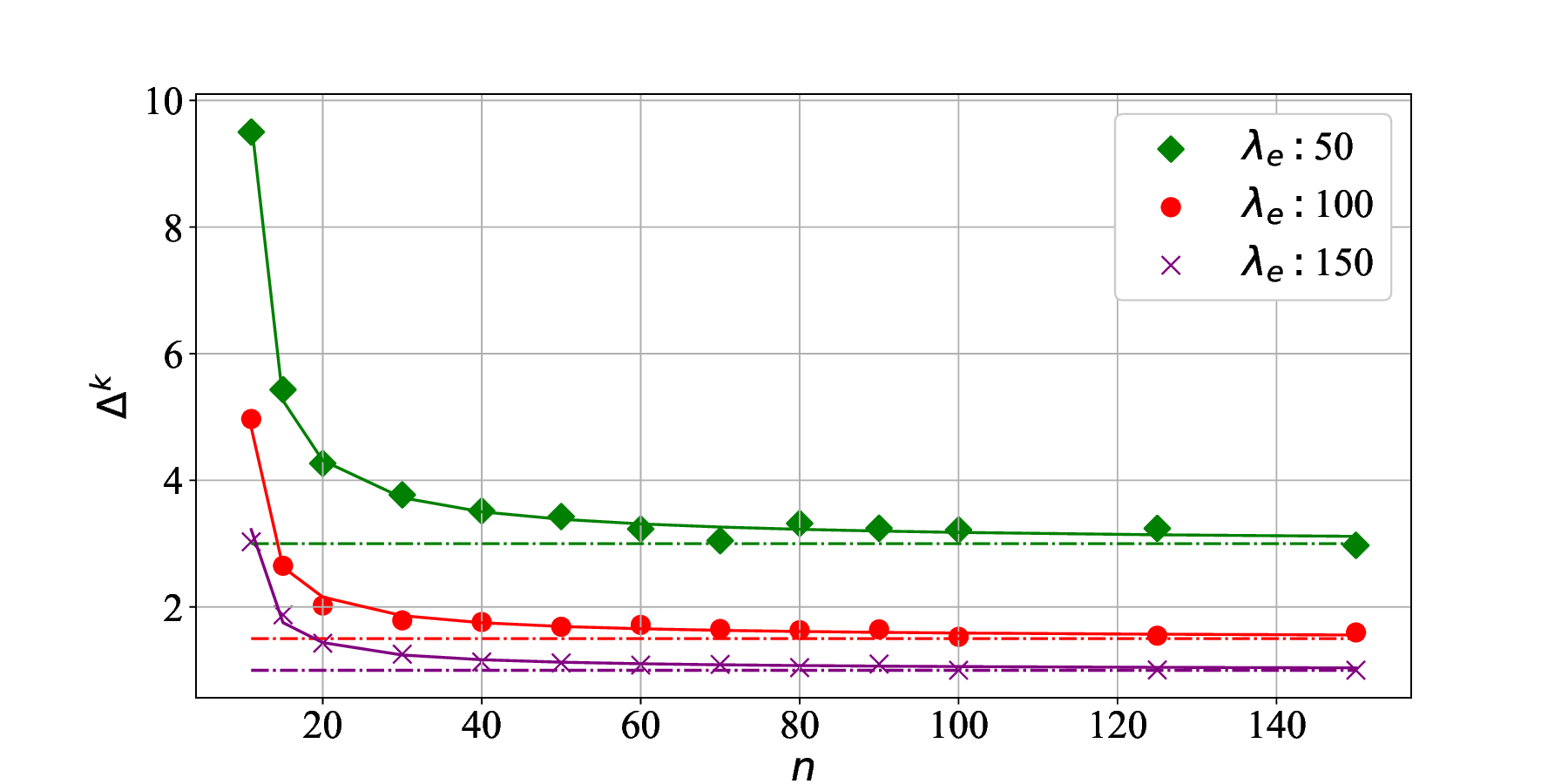}  \vspace{-3.5mm}
    \caption{$\Delta^k$ as a function of $n$ when $k=10$ and $\lambda_s=15$. Solid lines show the theoretical $\Delta^k$ while simulation results for $\lambda_e=\{50,100,150\}$ selections are marked by $\blacklozenge,\bullet,\times$, respectively. Dashed lines show the theoretical asymptotic value of $\Delta^k$ on $n$.}\vspace{-5mm}
    \label{fig:asym_on_on}
\end{figure}

\begin{figure}
	\begin{center}
		\subfigure[]{%
			\includegraphics[width=0.45\linewidth]{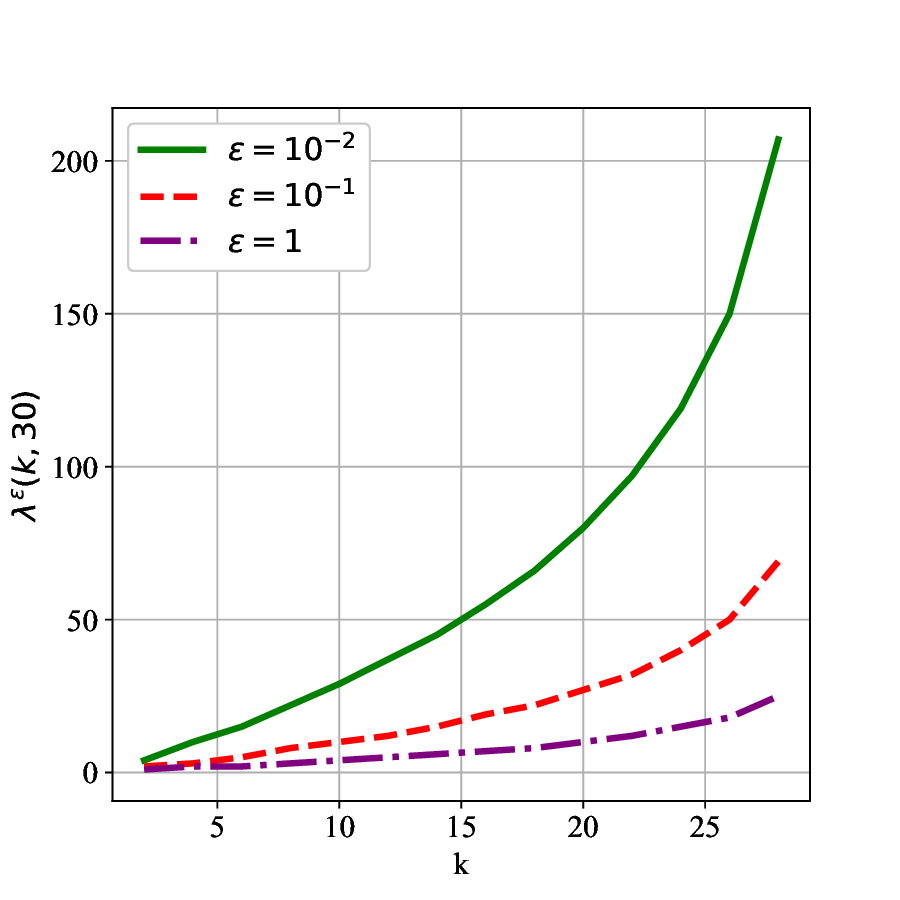}} \hspace{-1mm}
		\subfigure[]{%
			\includegraphics[width=0.45\linewidth]{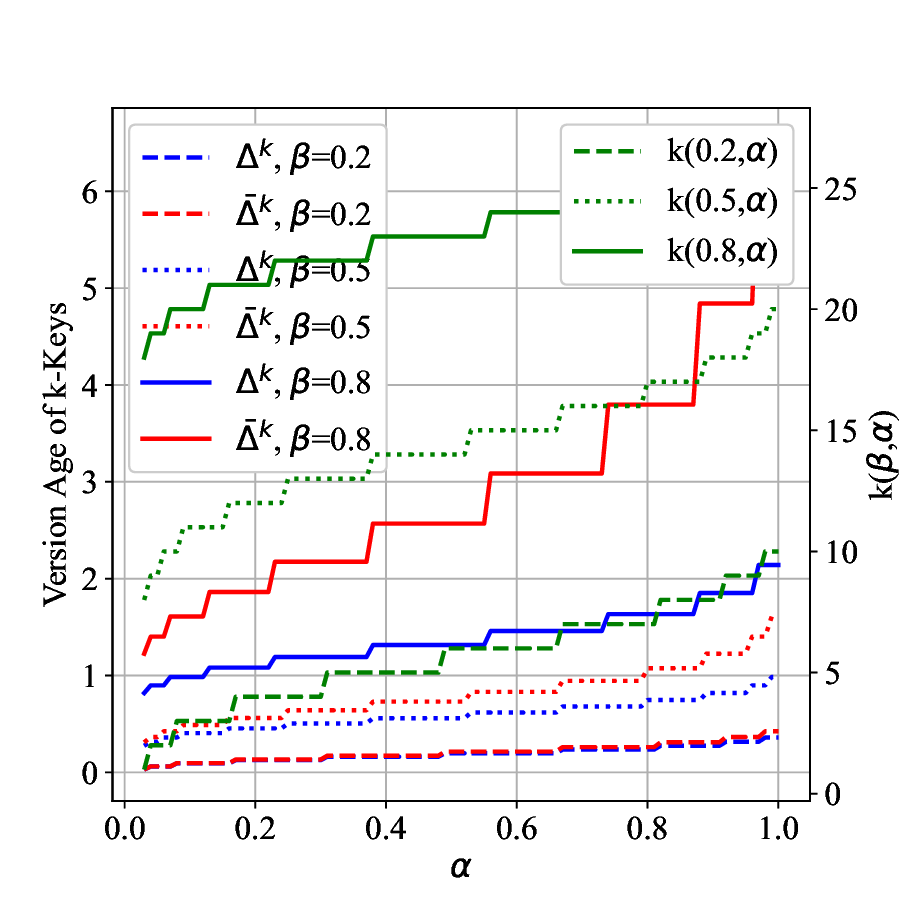}}
	\end{center}
 \vspace{-6mm}
	\caption{(a) The memory critical gossip rate $\lambda^\varepsilon(k,30)$ as a function of $k$ for $\varepsilon \in \{10^{-2},10^{-1},1\}$ and (b) $\bar{\Delta}^k$,${\Delta}^k$ and $k(\beta,\alpha)$ as a function of $\alpha \in[0,1]$ for $\beta\in \{0.2,0.5,0.8\}$ when $\lambda_s=15$ and $n=30$.}
	\label{fig:lambda_epsilon}
	\vspace{-5mm}
\end{figure}

\vspace{-2mm}
In this section, we compare empirical results obtained from simulations to our analytical results.
\begin{figure}
	\begin{center}
		\subfigure[]{%
			\includegraphics[width=0.47\linewidth]{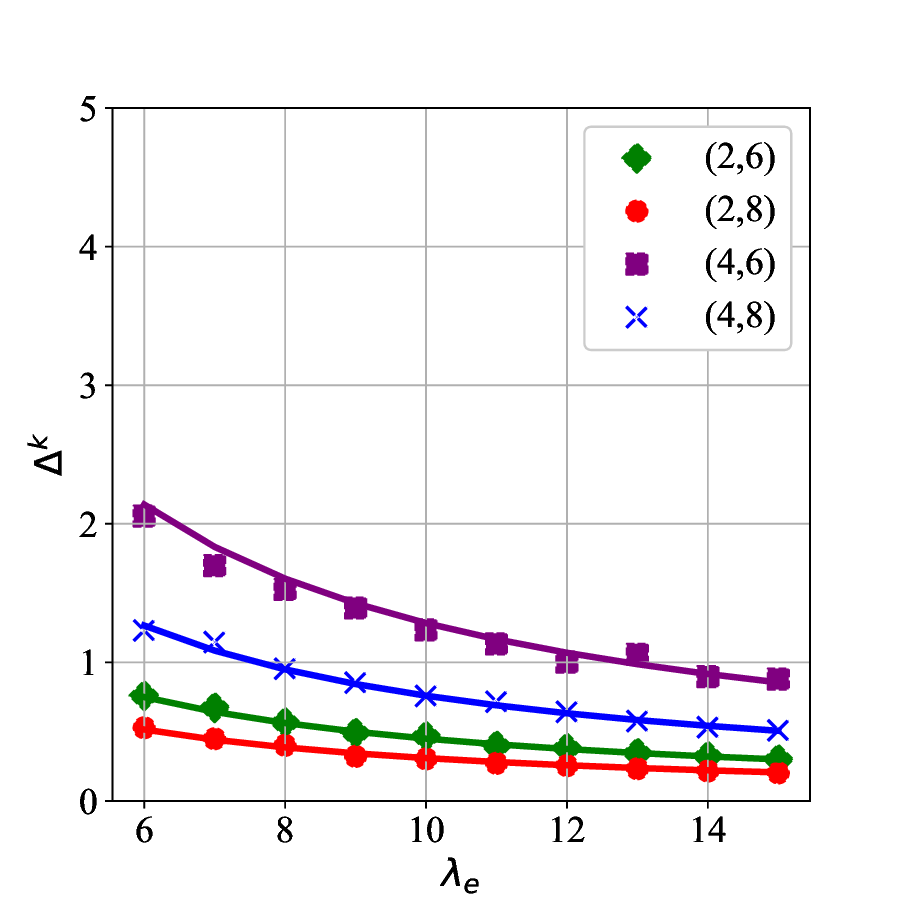}} \hspace{-1mm}
		\subfigure[]{%
			\includegraphics[width=0.47\linewidth]{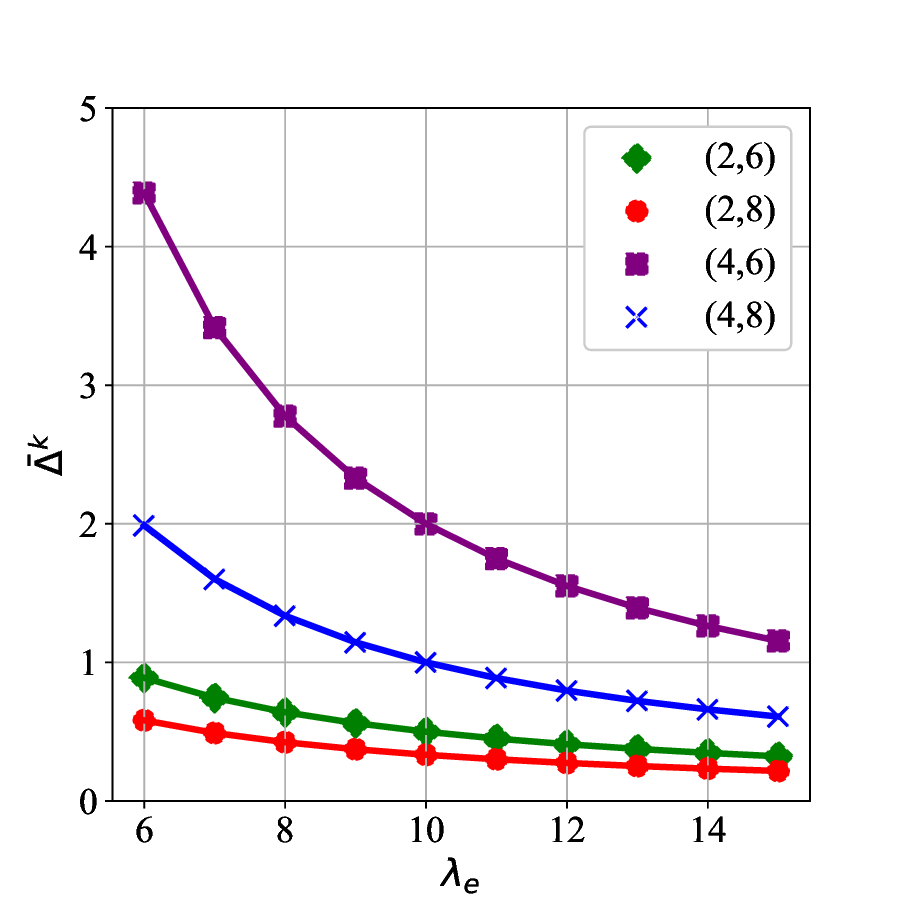}}
	\end{center}
 \vspace{-6mm}
	\caption{(a) $\Delta^k$ and (b) $\bar{\Delta}^k$ as a function of $\lambda_e$ on a SHN when $\lambda_s=10$. Solid lines in Fig.~\ref{fig:theoric_experimental}(a) and Fig.~\ref{fig:theoric_experimental}(b) show theoretical $\Delta^k$ and theoretical $\bar{\Delta}^k$, respectively. Simulation results for $(2,6)$,$(2,8)$,$(4,6)$,$(4,8)$ TSS are marked by $\blacklozenge,\bullet,\blacksquare,\times$, respectively.}
	\label{fig:theoric_experimental}
	\vspace{-5mm}
\end{figure}
We consider a SHN. Fig.~\ref{fig:theoric_experimental} depicts the simulation and the theoretical results for both $\Delta^k$ and $\bar{\Delta}^k$ as a function of gossip rate $\lambda_e$ when $\lambda_s=10$. The simulation results for $\Delta^k$ and $\bar{\Delta}^k$ align closely with the theoretical calculations provided in Theorems~\ref{thm:hetero_w_memory_age} and \ref{thm:hetero_wo_memory_age}, respectively. In both schemes, we observe that the version age of $k$-keys for a node decreases with the rise in the gossip rate $\lambda_e$, while keeping the network size $n$ and $\lambda_s$ constant. 
We observe, in Fig.~\ref{fig:theoric_experimental}(a) and Fig.~\ref{fig:theoric_experimental}(b), that both $\Delta^k$ and $\bar{\Delta}^k$ increase with the growth of $k$ for a fixed $\lambda_e$ and $n$ and they decreases as $n$ increases for fixed $\lambda_e$ and $k$. Also, we observe, in Fig.~\ref{fig:theoric_experimental}, that $\Delta^k$ is {\em less than} $\bar{\Delta}^k$ for the same values of $\lambda_e,\lambda_s$ and $(k,n)$.  Fig.~\ref{fig:asym_on_on} depicts $\Delta^k$ as a function of the number of the nodes $n$ for various gossip rates $\lambda_e$. We observe, in Fig.~\ref{fig:asym_on_on}, that $\Delta^k$ converges to $\frac{k\lambda_s}{\lambda_e}$ as $n$ grows. It aligns with Cor.~\ref{cor:w_memory_scale}. 
Fig.~\ref{fig:lambda_epsilon}(b) depicts $ \bar{\Delta}^k $ and $ \Delta^k $ as functions of $ (\beta, \alpha) $ for the given $D(k,n,\beta)$ in the introduction. We observe that in Fig.~\ref{fig:lambda_epsilon}(b), both $ \bar{\Delta}^k $ and $ \Delta^k $ increase as the required precision $ \alpha $ increases. Additionally, for the same precision rate $ \alpha $, higher noise in the measurement $ \beta $ results in greater $ \bar{\Delta}^k $ and $ \Delta^k $.

To quantify the value of memory in a network, we define {\em the memory critical gossip rate} of a $(k,n)$-TSS network for a margin $\varepsilon$, denoted by $\lambda^\varepsilon(k,n)$, as the smallest gossip rate $\lambda_e$ such that $|\Delta^k-\bar{\Delta}^k|\leq \varepsilon$. We first observe, in Fig.~\ref{fig:lambda_epsilon}(a), that $\lambda^\varepsilon(k,n)$ xponentially increases as $k$ increases for fixed $n$. Now, consider the event $E\!:=\!\{\mathcal{X}_{(k:n-1)}\leq U\}$, one can easily see that $Pr(E)$ converges to $1$ as $\lambda_e$ increases (frequent gossipping between nodes) or $k$ decreases for fixed $n$. In this case, the expectation $\mathbb{E}[\min(\mathcal{X}_{(k:n-1)},U)]\!\to\!\mathbb{E}[\mathcal{X}_{(k:n-1)}]$ in~\eqref{eqn:hetero_wo_memory_age}. It implies that $\bar{\Delta}^k$ approach to $\Delta^k$ as $Pr(E)$ goes to $1$. These observations show that $\bar{\Delta}^k$ approaches ${\Delta}^k$ as $\lambda_e$ increases or $k$ decreases.






\noindent {\bf Conclusion.} In this work, we have provided closed-form expressions for the version age of $k$-keys for both with memory and memoryless schemes. In our work, nodes only send the keys that are received from the source node, ensuring that any set of messages on the channels is not sufficient to decrypt the message at any time. An alternative approach might be to consider the case where nodes can share keys that they received from other nodes.

\enlargethispage{-1.2cm} 
 
\bibliographystyle{IEEEtran}
\bibliography{aoi.bib}

\end{document}